\documentclass
[preprint,prl,showpacs,noshowkeys,10pt,twocolumn,tightenlines,letterpaper,lengthcheck]{revtex4}%
\usepackage{amsfonts}
\usepackage{amsmath}
\usepackage{amssymb}
\usepackage[dvips]{graphicx}%
\begin{document}
\preprint{ }
\title{Suppression of inelastic electron-electron scattering in Anderson Insulators}
\author{Z. Ovadyahu}
\affiliation{Racah Institute of Physics, The Hebrew University, Jerusalem 91904, Israel }

\begin{abstract}
We report on measurements of absorption from applied ac fields in
Anderson-localized indium-oxide films. The absorption shows a roll-off at a
frequency that is much smaller than the electron-electron scattering rate
measured at the same temperature in diffusive samples of this material. These
results are interpreted as evidence for discreteness of the energy spectrum.

\end{abstract}
\maketitle

Inelastic scatterings of electrons play an important role in the properties of
Fermi gas systems. The most frequently encountered types of such events are
electron-electron (e-e) and electron-phonon (e-ph) scatterings.
Energy-exchange via efficient e-e scattering is vital for establishing the
Fermi-Dirac distribution, which defines the electron temperature. The
electron-phonon inelastic collisions is needed to maintain steady-state
situations, and in particular, are responsible for the validity of Ohms law.
Either process may lead to de-coherence of the electrons and thus control the
quantum effects exhibited by the system.

Both, the e-e inelastic-rate $\gamma_{in}^{e-e},$ and the e-ph inelastic-rate
$\gamma_{in}^{e-ph}$, are temperature dependent. The specific form of these
rates (typically a power-law of the temperature $T)$, depend on system
dimensionality, temperature range, and the type and degree of disorder. In the
thermodynamic limit of diffusive systems however, and at low, yet
experimentally accessible, temperatures, $\gamma_{in}^{e-e}$ is usually larger
than $\gamma_{in}^{e-ph}$ \cite{1}.

In this Letter we report on measurements of the energy absorbed by the
charge-carriers in Anderson localized indium-oxide films from electric fields
as function of frequency $f$. The technique used in this work utilizes a
unique property of electron-glasses \cite{2}; the excess-conductance produced
by a non-Ohmic field reflects the energy absorbed by the charge carriers. This
technique allows a measurement on systems with very small volume, sensitive
enough to allow for weak absorption from electric fields, and most important
for this work - can be carried over a large frequency range.

The measurements described below suggest that $\gamma_{in}^{e-e}$ in the
electron-glass is dramatically suppressed relative to its value at the
diffusive regime of the same material, and thermalization of the electronic
system is presumably governed by $\gamma_{in}^{e-ph}$. These results may be
relevant for testing many-body localization models \cite{3,4}.

The electron-glass samples used in this study were thin films of crystalline
indium-oxide (In$_{\text{2}}$O$_{\text{3-x}}$) e-gun evaporated on glass
substrates. Lateral size of the samples used here were $L=$2~mm long and
$W=$1~mm wide. Their thickness ($d=$52~\AA )~and stoichiometry (fine-tuned by
UV-treatment \cite{5}) were chosen such that at the measurement temperatures,
the samples had sheet resistance R$_{\mathbf{\square}}$ in the range
6~M$\Omega$-12~M$\Omega$. Conductivity of the samples was measured using a two
terminal ac technique employing a 1211-ITHACO current preamplifier and a
PAR-124A lock-in amplifier. The measurements were performed with the samples
immersed in liquid helium at $T$=4.11~K held by a 100 liters storage-dewar.
This made it possible to perform measurements of samples while maintaining a
stable bath-temperature over the very long period required for the type of
experiments described below. Fuller details of sample preparation,
characterization, and measurements techniques are given elsewhere \cite{6,7}.

Several sources were used for exciting the system by non-Ohmic fields; the
internal oscillator of the PAR124A (up to 2~kHz and 10~Vrms) (Fluke PM5138A
(dc and up to 10~MHz and 40~Vpp), and Tabor WS8101 (up to 100~MHz and 16~Vpp).

The main technique used in this study is the `stress-protocol' previously used
in aging experiments \cite{8}. The procedure is composed of the following
stages: After the sample is equilibrated at the measuring temperature
(typically for 20 hours), its conductance ~$G$ versus time is recorded while
keeping the electric field $F$=$F_{0}$ small enough to be as close to the
Ohmic regime as possible. This defines a baseline `equilibrium' conductance
$G(F_{0})$. Next, $F$ is switched to a non-Ohmic $F_{s}$ which causes the
conductance to increase by a predetermined $\Delta G(0)$. $F_{s}$ is kept on
the sample for a time $t_{w}$, then the field is switched back to $F_{0}$ and
the conductance is continued to be measured for $\approx$5$\cdot t_{w}$. This
last stage is depicted in figure 1 as a relaxation of $G$ towards the
equilibrium $G(F_{0})$ with a logarithmic law characteristic of relaxation
processes of electron-glasses \cite{2}. A measure of the magnitude of the
excess conductance that results from the stress is $\delta G_{rel}$, defined
by extrapolating $\delta G(t)$ to 1~second as illustrated in the inset to Fig.
1. $\delta G(t)$ is $G(F_{0},t)-G(F_{0})$ where the origin of time $t$ is
taken as $t_{w}$+1 (i.e., 1 second after $F_{s}$ is reset to $F_{0}$).
\begin{figure}
[ptb]
\begin{center}
\includegraphics[
trim=0.147790in 0.652868in 0.260393in 0.563401in,
height=3.1972in,
width=3.4601in
]%
{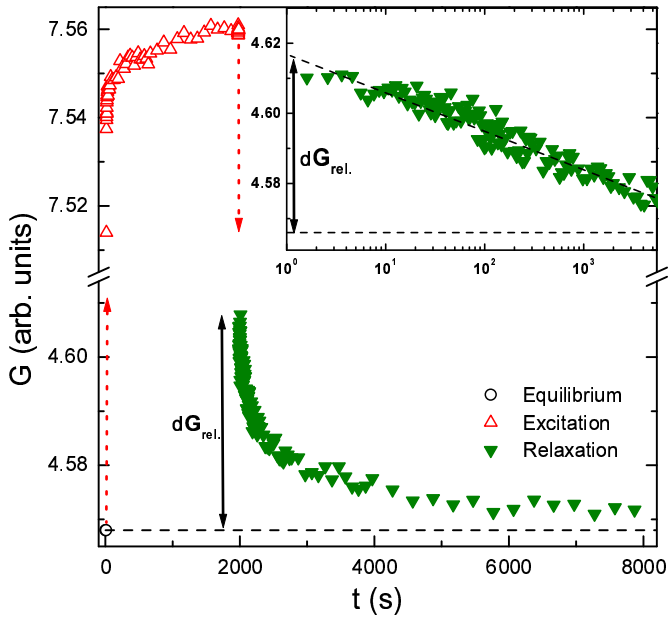}%
\caption{Conductance versus time $G(t)$ illustrating a typical
`stress-protocol' run. The sample is In$_{\text{2}}$O$_{\text{3-x}}$ film with
an equilibrium R$_{\square}$=12.5~M$\Omega$ measured at $T$=4.1K with $F_{0}%
$=5~V/m and $F_{s}$=500~V/m both at 730~Hz. The inset shows the logarithmic
relaxation of $\delta G(t)$ and the definition of $\delta G_{rel}$ (upper
plot). Dashed lines delineate the equilibrium conductance $G(F_{0})$.}%
\end{center}
\end{figure}

The relaxation of $\delta G(t)$ reflects the release of excess energy
accumulated during $t_{w}$ through Joule-heating produced by the field; While
$F_{s}$ is on, the energy absorbed by the electronic system gives rise to an
excess phonons within the sample, making it somewhat `hotter' than the bath. A
steady-state is established by the flow of energy from the sample-phonons into
the thermal-bath. The increased density of high energy phonons (over the
phonon population in equilibrium at the bath temperature), randomizes the
charge configuration of the electron-glass in a similar vein that raising the
bath-temperature would \cite{9}. This is reflected in the sluggish conductance
increase during $t_{w},~$and to the ensuing $\delta G(t)$ after the stress is
relieved (Fig.1). $\delta G_{rel}$ is used here as a \textit{relative} measure
of the energy absorbed by the electronic system from the stress field (i.e.,
no quantitative value is assigned to this measure, which, as will be clear
below, does not affect our conclusions).

The intriguing finding of this work is the non-trivial dependence of $\delta
G_{rel}$ on the frequency of the stress field. In these experiments, the
stress-protocol is repeated using fields $F_{s}(f)$ of different frequencies
$f$ as described next.

For a meaningful comparison between results of different frequencies,
$F_{s}(f)$ is applied for the same $t_{w}$ and its amplitude chosen such as to
affect the \textit{same} $\Delta G(0)$ for each $f$. This procedure then
requires a pre-knowledge of the $F_{s}$ amplitude that achieves the target
$\Delta G(0)$. This is obtained from plots of $G(F)$ taken independently at
each frequency to be measured. Examples of $G(F)$ plots measured at different
frequencies are shown in figure 2.%
\begin{figure}
[ptb]
\begin{center}
\includegraphics[
trim=0.219731in 0.044331in 0.272122in 0.091885in,
height=3.6409in,
width=3.3693in
]%
{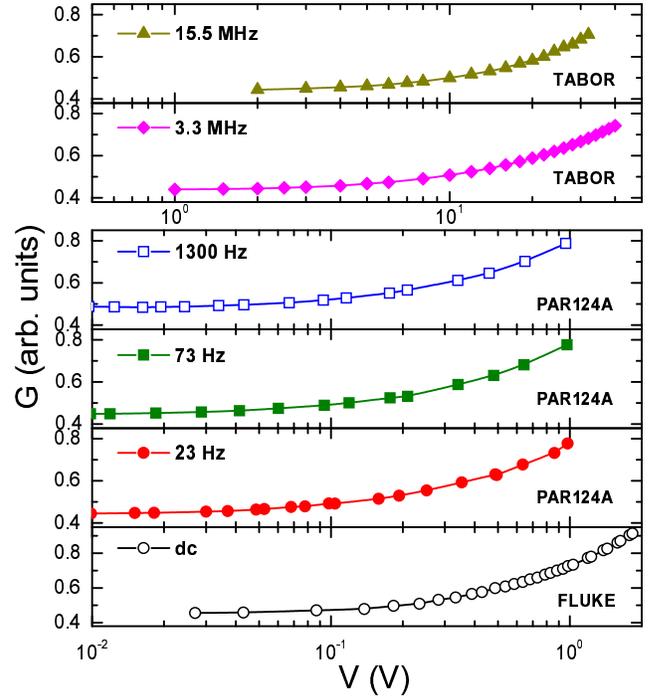}%
\caption{The dependence of the sample conductance on the applied voltage
measured at different frequencies. The instrument used as the voltage source
for a given plot is marked on the bottom right corner.}%
\end{center}
\end{figure}

These plots were used in the first series of the stress-protocol experiments
where $\Delta G(0)/G(F_{0})$ was set to be 0.75$\pm$0.2 at all frequencies
using a sample with Ohmic conduction (per-square) $G(F_{0})$=8.33$\cdot
$10$^{-8}~\Omega^{-1}$. Note that the functional dependence of these $G(V)$
curves is only weakly dependent on the frequency of the measurement. The
$\delta G_{rel}$ that results from the stress protocol at these frequencies,
on the other hand, \textit{is} \textit{frequency dependent} (figure 3): It is
essentially constant (actually, starting from dc which cannot be explicitly
shown on Fig.~3) up to some frequency, then it gradually rolls-off, and tends
to vanish for $f\geq$10$^{8}~$Hz. Note that this is consistent with the lack
of absorption at the microwaves frequencies reported previously \cite{7}.

Also shown in Fig.~3 are the respective results of another sample with Ohmic
conduction (per-square at 4.1K) $G(F_{0})$=1.64$\cdot$10$^{-7}~\Omega^{-1}%
$.$~$This sample was obtained from the first by UV-treatment, which amounts to
a small change of stoichiometry \cite{5}. The two samples have therefore
identical geometry, and crystallites-size. They differ in their $G(F_{0}),$ in
their localization length $\xi$ (c.f., Fig.~4 below), and in that $\Delta
$G(0)/G$_{0}$ was set to be 0.65$\pm$0.2 for the second sample. To achieve the
targeted $\Delta$G(0)/G$_{0},$ the applied voltages during $t_{w}$ (1400$\pm
$10$~\sec$ for both samples), were 1~V(rms) and 0.5~V(rms) for the first and
second samples respectively.

The linearity of $\delta G_{rel}$ with $\Delta G(0)$ was ascertained by
measuring $\delta G_{rel}$ at half and twice of the nominal $\Delta G(0)$ of
the series and for both: $f$=23~Hz and $f$=0.8~MHz. It should also be noted
that to achieve a constant $\Delta G(0)$ higher frequencies the field
amplitudes across the sample were, if anything, somewhat larger than at low
frequencies. Therefore the vanishing values of $\delta G_{rel}$ at high
frequencies cannot be due to spurious effects; the sample and its environment
(substrate etc.) are the same at all stress frequencies. Also, at the higher
frequencies, where of $\delta G_{rel}$ is diminishing with $f$, the electrons
are subjected to as large or even larger field amplitudes than used at low
frequencies to get the same $\Delta G(0)$.

The roll-off frequency of $\delta G_{rel}(f)$ turns out to be of the order of
the rate by which energy acquired from the field is dissipated into the bath
(presumably, $\gamma_{in}^{e-ph}$); Under a bias voltage $V$ (dc or low enough
$f$), acting on resistance $R(V),$ the following expression may be used to
estimate the rate of dissipated energy in a steady state:%
\begin{equation}
\frac{V^{\text{2}}}{R(V)}=C_{el}(T)(\Delta T)\tau_{in}^{\text{-1}} \label{1}%
\end{equation}

Here $C_{el}$ is the electronic heat-capacity, $\Delta T$ is the temperature
difference between the electronic system and the phonons, and $\tau
_{in}^{\text{-1}}$ is the rate of energy dissipation. We shall use the
free-electron expression: $\frac{\text{2}}{\text{3}}\pi^{\text{2}%
}N(0)k_{\text{B}}^{\text{2}}Tu$ for $C_{el}(T),$ where$~N(0)\simeq$2$\cdot
$10$^{\text{45}}~$J$^{\text{-1}}$m$^{\text{-3}}$ is the thermodynamic density
of states at the Fermi energy for In$_{\text{2}}$O$_{\text{3-x}}$[5], and
$u=LWd$ is the sample volume ($u=$ 1.04$\cdot10^{\text{-14}}$~m$^{\text{3}}$
for our samples). To get a rough estimate for $\Delta T$ we neglect the
difference in temperature between the system and the phonons of the thermal
bath, and assume that the conductance of the sample versus temperature
(measured under near-Ohmic conditions), reflects the electron temperature
\cite{10}. The temperature dependence of $G_{0}$ is shown in figure 4 for the
two samples. Using these data gives for $\Delta T$=0.71K for the first sample,
and $\Delta T$=0.74K for the second sample, which from Eq.~1 we get $\tau
_{in}^{-1}$=3.5$\cdot$10$^{\text{5}}$ and 7.1$\cdot$10$^{\text{5}}$ Hz
respectively. These rates are somewhat under-estimated for two reasons;
disregarding the elevated temperature of the system-phonons, and assuming that
the only source for $\Delta G(0)$ is `heating' (i.e., neglecting the
field-assisted hopping contribution \cite{11}). These omissions overestimate
$\Delta T$ and thus lead to an underestimated rates by about an order of
magnitude \cite{12}, still within the region where $\delta G_{rel}$ decays
which extends over nearly four decades, presumably reflecting the distribution
of $\gamma_{in}^{e-ph}$. By comparison, the excess conductance in these
electron-glasses has been observed to relax over at least six decades in time,
and the logarithmic law of this relaxation is suggestive of a wide
rate-distribution. Wide rate-distributions are ubiquitous in condensed matter
systems, and invariably the underlying reason is disorder. This is a natural
occurrence in hopping conductivity; in a realistic system the localization
length $\xi$ is distributed over a range determined by the strength (and type)
of the quenched disorder. Electronic transition probabilities depend on
wave-function overlap, and even on a single-particle level, a mildly wide
$\xi$-distribution may translate into an exponentially wide distribution. This
complication may be ignored when dealing with the macroscopic conductance due
to its percolative nature but it should be evident in energy relaxation
processes.
\begin{figure}
[ptb]
\begin{center}
\includegraphics[
trim=0.230519in 0.444153in 0.384467in 0.222859in,
height=3.2603in,
width=3.3909in
]%
{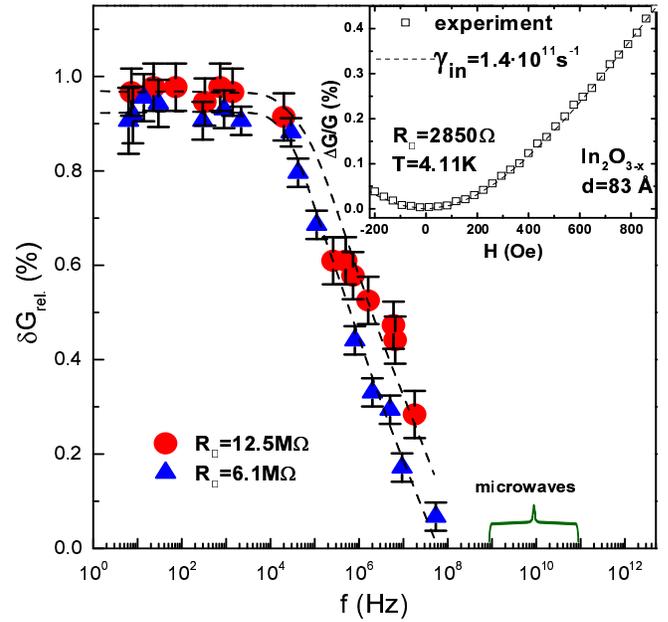}%
\caption{The frequency dependence of the amplitude of excess conductance
following the stress protocol for the two studied samples (see text for
details). Dashed lines are guides for the eye. The inset depicts a
magneto-conductance measurement for a diffusive film of the same material
fitted to the weak-localization theory of Ref. \cite{10}.}%
\end{center}
\end{figure}

The inability of the electrons to absorb energy from the field is expected
when the drive frequency exceeds $\gamma_{in}^{\max},$ the maximum rate of
their energy-exchange with any available sub-system. In the diffusive
regime,$\gamma_{in}^{\max}$ is the electron-electron inelastic rate
$\gamma_{in}^{e-e}$ which at low temperatures, is much larger than the
electron-phonon inelastic rate $\gamma_{in}^{e-ph}$. This, in particular,
holds true for In$_{\text{2}}$O$_{\text{3-x}}~$that has uncommonly high Debye
temperature (1050-1100~K \cite{13}) and thus relatively low $\gamma
_{in}^{e-ph}$ in line with the above estimate of $\approx$10$^{\text{6}}%
\sec^{-1}$ at $\approxeq$4~K. The e-e inelastic rate in a two-dimensional (2D)
In$_{\text{2}}$O$_{\text{3-x}}~$film with $R_{\square}\approx$10$^{\text{3}%
}~\Omega$ at $\approx$4K is of the order of $\gamma_{in}^{e-e}\approx
$10$^{\text{11}}\sec^{-1}$. This is based on magneto-conductance measurements
and a fit to the weak-localization theory \cite{14}. An example is shown in
the inset to Fig.~3. The dephasing rate $\gamma_{in}\approx$1.4$\cdot
$10$^{\text{11}}\sec^{\text{-1}}$, that results from the fit to the theory, is
consistent with the Abrahams et al model for e-e inelastic rate \cite{15}. In
the diffusive regime and for thin films the theory anticipates that
$\gamma_{in}^{e-e}$ should only \textit{increase} with disorder . Our
experiments suggest that, for sufficiently strong disorder, this trend has
been reversed; $\gamma_{in}^{e-e}$ in both films is smaller than its value in
the $R_{\square}\approx$3~k$\Omega$ sample by \textit{at least} three orders
of magnitude: Note that the technique used here can only put an upper bound on
the inelastic rate; it cannot determine the actual $\gamma_{in}^{e-e},$ which
might in fact be much smaller than $\gamma_{in}^{\max}$ estimated from the
roll-off in Fig.~3.
\begin{figure}
[ptb]
\begin{center}
\includegraphics[
trim=0.172031in 1.213851in 0.254137in 0.760067in,
height=2.7951in,
width=3.3849in
]%
{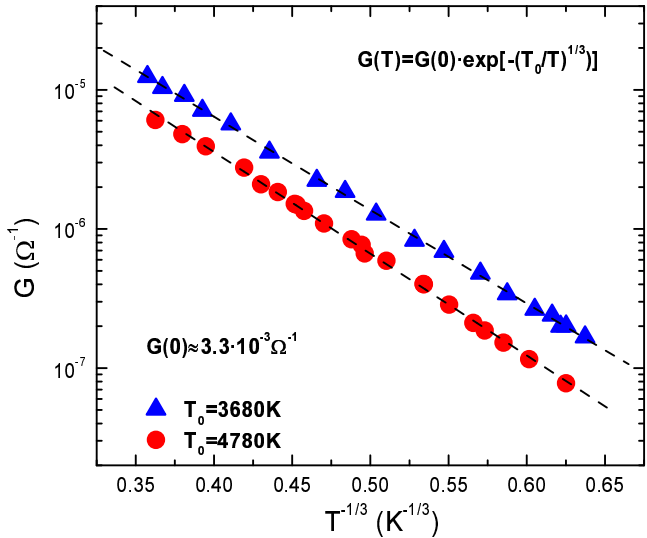}%
\caption{Temperature dependence of the conductance of the samples used in this
work and their relevant parameters derived from the plots.}%
\end{center}
\end{figure}

This dramatic suppression in $\gamma_{in}^{e-e}~$of the same material at
$\approx$4~K is presumably related to the change in the transport mechanism,
which in 2D In$_{\text{2}}$O$_{\text{3-x}}~$films with $R_{\square
}\eqslantless h/e^{\text{2}}~$is diffusive while sample with $R_{\square}\gg
h/e^{\text{2}}$ exhibit variable-range-hopping (VRH). The samples studied here
obey Mott's VRH in 2D, given by $G(T)=G(0)\cdot\exp\{(T_{0}/T)^{\text{1/3}}\}$
where T$_{0}\approx$[$k_{B}N(0)\xi^{\text{2}}d$]$^{\text{-1}}$ (Fig.~4). Over
the temperature range relevant for our experiments the thermal energy is
\textit{much} smaller than the mean level-spacing $\delta=$[$N(0)\xi
^{\text{2}}d$]$^{\text{-1}}$. In a non-interacting picture, the inherent
\textit{discreteness} of the energy-spectrum of the system should be evident
in the transport when $R_{\square}\gg h/e^{\text{2}}$ but it is washed-out due
to life-time broadening in the limit $R_{\square}\eqslantless h/e^{\text{2}}$
where the inelastic diffusion length is smaller than $\xi$. Discreteness is
indeed the key to the orders-of-magnitude change in $\gamma_{in}^{e-e},$ and
this may be understood in the following way:

In the simplest scenario, of independent set of electronic states localized
over regions with spatial extent $\xi,$ the problem is reduced to that of a
quantum dot with volume $\xi^{\text{2}}d$. This case has been studied
\cite{16} with the result that $\gamma_{in}^{e-e}$ is exponentially suppressed
once $k_{B}T\leq$ $\delta.$ For our samples $\delta$ is about three orders of
magnitude$~$larger than the temperature of the measurements, and~therefore,
$\gamma_{in}^{e-e}$ should be vanishingly small. Energy exchange between
electrons and their thermalization hinge on the existence of another
sub-system with a continuous spectrum, e.g., the phonons bath. This
single-particle scenario seems to be consistent with our data.

However, an Anderson insulator is not an independent set of quantum dots, and
Coulomb interactions cannot be ignored in such a disordered medium with its
impaired screening. The samples studied here are electron-glasses, and
electron-electron interactions manifestly play a crucial role in their
transport properties \cite{17}.

A fundamental question in this context is whether many-body excitations could
de-localize the system and render the energy spectrum continuous. These issues
were considered by Anderson \cite{18}, Fleishman and Anderson \cite{19}, and
recently by Gornyi et al \cite{3}, and Basko et al \cite{4}. For short range
interactions (and in a 2D system, lacking a mobility edge) the spectrum is
expected to be discrete and thus, at low temperatures, electron can exchange
energy among themselves only by virtue of the phonons bath (or another
sub-system with a continuous spectrum). In other words, the rate of energy
exchange between electrons is effectively limited by the electron-phonon
inelastic rate $\gamma_{in}^{e-ph}$. Our results are also consistent with
this, many-body scenario, which is physically more relevant for electron-glasses.

A corollary of many-body-localization is that, at low temperatures, the energy
$\Delta\varepsilon_{ij}~$needed for an electron to hop from localized site $i$
to localized site $j,$ must be supplied by a continuous bath. Hopping models
traditionally assumed that it is the electron-phonon interaction that is
involved. However the magnitude of the pre-exponential factor of $G(T)$ (see,
e.g., Fig.~4) and its weak temperature dependence have yet to be accounted for
by a model based on a phonon-bath. Identifying the nature of the bath that
meets the constraints implied by the many experiments published over the last
four decades, as well as the one presented here, is a challenge for theory.

This research was supported by a grant administered by the US Israel
Binational Science Foundation and by the Israeli Foundation for Sciences and Humanities.


\begin{thebibliography}{99}                                                                                               %


\bibitem {1}A. Mittal, \textit{Quantum Transport in Submicron Structures},
Advanced NATO Proceedings Kluwer Academic, Dordrecht, (1996).

\bibitem {2}J. H. Davies et al, Phys. Rev. Lett, \textbf{49}, 758 (1982); M.
Gr\"{u}newald B. Pohlman, L. Schweitzer, and D. W\"{u}rtz,, J. Phys. C,
\textbf{15}, L1153 (1982); M. Pollak and M. Ortu\~{n}o, Sol. Energy Mater.,
\textbf{8}, 81 (1982); M. Pollak, Phil. Mag. B\textbf{ 50}, 265 (1984); G.
Vignale, Phys. Rev. B\textbf{ 36}, 8192 (1987); M. M\"{u}ller and L. B. Ioffe,
Phys. Rev. Lett. \textbf{93}, 256403 (2004); C. C. Yu, Phys. Rev. Lett.,
\textbf{82}, 4074 (1999); Vikas Malik and Deepak Kumar, Phys. Rev. B
\textbf{69}, 153103 (2004); D. R. Grempel, Europhys. Lett., \textbf{66,} 854
(2004); Eran Lebanon, and Markus M\"{u}ller, Phys. Rev. B\textbf{\ 72}, 174202
(2005); A. Amir, Y. Oreg, and Y. Imry, Annu. Rev. Condens. Matter Phys.
\textbf{2,} 235 (2011).

\bibitem {3}I. V. Gornyi, A. D. Mirlin, and D. G. Polyakov, Phys. Rev. Lett.
\textbf{95}, 206603 (2005).

\bibitem {4}D. M. Basko, I. L. Aleiner, and B. L. Altshuler, Ann. Phys. (N.Y.)
\textbf{321}, 1126 (2006).

\bibitem {5}Z. Ovadyahu, J. Phys. C: Solid State Phys., \textbf{19}, 5187
(1986); V. Chandrasekhar and R. A. Webb, J. Low Temp. Phys. \textbf{97}, 9 (1994).

\bibitem {6}A. Vaknin, Z. Ovadyahu, and M. Pollak, Phys. Rev. B\textbf{ 65},
134208 (2002).

\bibitem {7}V. Orlyanchik A Vaknin, and Z. Ovadyahu, Phys. Stat. Sol.,
B\textbf{ 230}, 67 (2002); V. Orlyanchik, and Z. Ovadyahu, Phys. Rev. Lett.,
\textbf{92}, 066801 (2004).

\bibitem {8}Z. Ovadyahu, Phys. Rev. B \textbf{83}, 235126 (2011).

\bibitem {9}Z. Ovadyahu, Phys. Rev. Lett., \textbf{102}, 206601 (2009).

\bibitem {10}S. Marnieros, L. Berge, A. Juillard, and L. Dumoulin, Phys. Rev.
Lett. \textbf{84}, 2469 (2000).

\bibitem {11}R. M. Hill, Philos. Mag. \textbf{24}, 1307 (1971); N. Apsley and
H. P. Hughes, Philos. Mag. \textbf{31}, 1327 (1975); B. I. Shklovskii, Fiz.
Tekh. Poluprovodn. \textbf{6}, 2335 (1972) [Sov.Phys. Semicond. \textbf{6},
1964 (1973)]; M. Pollak and I. Riess, J. Phys. \textbf{C9}, 2339 (1976).

\bibitem {12}This estimate is based on comparison methods used in: Z. Ovadyahu
Phys. Rev. B \textbf{73}, 214208 (2006).

\bibitem {13}I. Schwartz S. Shaft, A. Moalem and Z. Ovadyahu, Phil. Mag. B
\textbf{50}, 221 (1984); X. D. Liu E. Y. Jiang, and D. X. Zhang, J. Appl.
Phys. \textbf{104}, 073711 (2008).

\bibitem {14}S. Hikami A. I. Larkin, and Y. Nagaoka, Prog. Theor. Phys. 63,
\textbf{707} (1980).

\bibitem {15}Elihu Abrahams P.W. Anderson, P. A. Lee, Phys. Rev. B
\textbf{24}, 6783 (1981).

\bibitem {16}Ya. M. Blanter, Phys. Rev. B \textbf{54}, 12807 (1996).

\bibitem {17}A. Vaknin Z. Ovadyahu, and M. Pollak, Phys. Rev. Lett.,
\textbf{81}, 669 (1998).

\bibitem {18}P. W. Anderson, Phys. Rev. \textbf{109}, 1492 (1958).

\bibitem {19}L. Fleishman and P. W. Anderson, Phys. Rev. B \textbf{21}, 2366 (1980).
\end{thebibliography}
\end{document}